\documentstyle[12pt,a4,rotate,epsf]{article}
\begin{document}
\begin{titlepage}
\headheight5pt
\headsep1pt
\topskip1pt
\setlength{\topmargin}{-1cm}
\hfill HD-THEP-94-08 \\
\vspace{3cm}
\begin{center}
\huge{BARYON ASYMMETRY FROM A TWO STAGE ELECTROWEAK PHASE TRANSITION ?\\
\vspace{1.5cm}}
\normalsize
A. Hammerschmitt\footnote{supported by Landes Graduierten F\"orderungs
Gesetz}\\
J. Kripfganz\footnote{supported by Deutsche Forschungsgemeinschaft}\\
M.G. Schmidt\vspace{0.8cm}\\
Institut f\"ur Theoretische Physik \\ der Universit\"at\\ Philosophenweg 16 \\
69120 Heidelberg\\ \vspace{1cm}
\end{center}
\begin{abstract}
%
%
%
We investigate an extension of the standard model with a second Higgs doublet,
showing a two stage phase transition. Wash-out of a baryon asymmetry after the
phase transition can be easily avoided in this class of models.
B+L transitions are more strongly suppressed  in the intermediate phase than
in the high temperature symmetric phase,however. Therefore, it becomes more
difficult if not impossible to generate a sufficient baryon asymmetry during
the phase transition.
\end{abstract}
\vfill
\end{titlepage}
\setlength{\topmargin}{-2cm}
\section{Introduction}
\setcounter{equation}{0}
Recently, significant progress has been made in understanding the
creation of a baryon number asymmetry of the universe. The electroweak
phase
transition \cite{Carrington},\cite{AndersonHall} is one particularly
attractive possibility for the generation of this asymmetry.
Indeed the standard model contains all the necessary ingredients
\cite{Sakharov}, i.e. baryon number violating interactions, C and CP
violation and a first order phase transition, that provides departure
from
thermal equilibrium. Unfortunately, problems arise when one tries to
quantitatively reproduce the observed baryon number asymmetry, using
the mechanism of \cite{CohenKaplanNelson},\cite{CoKaNe2}
in the framework of the standard model with the experimentally
allowed Higgs mass range  :

\begin{itemize}
\item{a baryon asymmetry created during the electroweak phase
transition is washed out in the true vacuum phase
\cite{Bochkarev},\cite{H&K&S}}
\item{the phase transition is not strongly first order
\cite{Juergen2},\cite{Carrington}}
\item{the CP violation due to the CKM matrix is small. It is under
controversial debate whether this suffices for baryon asymmetry
generation \cite{Shaposhnikov} or not  \cite{Orloff}}
\end{itemize}

Thus one has to face the situation that the standard model with one
Higgs might not be sufficient.

More complicated Higgs sectors, in particular models with two Higgs
doublets have been suggested as a possible way out.
In the case of two-doublet models additional CP violation
in the Higgs sector can be introduced and the enlarged number of
free parameters will help to solve the other problems.

In the present paper, we focus on
models having a two stage phase transition as proposed in
\cite{L&C}.  At some critical temperature
the universe would have first evolved towards some metastable
intermediate Higgs phase, followed by a first order phase
transition towards the present Higgs configuration
at some lower temperature. In this way one can easily avoid a
baryon wash-out in this last phase, just due to the lowering
of the transition temperature. The mass of the usual Higgs meson
(responsible for spontaneous symmetry breaking) can be as high as
a few hundred GeV, well above the current
experimental limit. The second Higgs doublet can also be quite
massive (again up to the order of 200 GeV) and would not have been
observed yet. The crucial problem for this model will be whether baryon
number violating processes in the intermediate phase
are fast enough to support baryon number generation.

In chapter 2 we introduce the model and discuss the appropriate
parameter region fora two-stage transition.
In chapter 3, B+L violation in the final phase is studied.
Chapter 4 contains the discussion of sphaleron transitions in
the intermediate phase. Finally we present our conclusions.

\section{The two Higgs effective potential}
\setcounter{equation}{0}
We will parametrize the potential of the  two  doublet model
in a slightly unconventional form following  \cite{L&C}

\begin{eqnarray}
V( \Phi_{1}, \Phi_{2} ) & = & 4 \lbrace k_{1} (\Phi_{1}^{
\dagger}\Phi_{1})^{2} +
k_{2}(\Phi_{2}^{\dagger} \Phi_{2})^{2} + k_{3} ( \Phi_{1}^{
\dagger}
\Phi_{1} ) (\Phi_{2}^{\dagger}\Phi_{2}) \nonumber\\
 & & \hspace{-20mm} +k_{4} \lbrack (\Phi_{1}^{\dagger}
\Phi_{1})(\Phi_{2}^{\dagger}\Phi_{2}) -
(\Phi_{1}^{\dagger}\Phi_{2}) (\Phi_{2}^{\dagger}\Phi_{1})
\rbrack +
\frac{k_{5}}{2}  \lbrack (\Phi_{1}^{\dagger}
\Phi_{2} )^{2} +  ( \Phi_{2}^{\dagger}\Phi_{1} )^{2}\rbrack
\\ & & - \mu_{1}^{2} \Phi_{1}^{\dagger} \Phi_{1} -
\mu_{2}^{2} \Phi_{2}^{\dagger} \Phi_{2} - \mu_{3}^{2}
Re(\Phi_{1}^{\dagger}
\Phi_{2}) - \mu_{4}^{2} Im(\Phi_{1}^{\dagger}\Phi_{2} )
\rbrace\nonumber
\end{eqnarray}

This version of the two Higgs potential can be easily
transformed
into the representation usually adopted
(e.g., \cite{HHGuide} ) but is more convenient for our
purpose.

We only couple the $\Phi_{1}$ doublet to the quarks to give
them a mass. This arrangement avoids tree level flavour
changing neutral currents.

$k_{5} $ is real and the corresponding term has a structure
which differs from
the one in \cite{L&C}.
This is not essential, however, since we ( as well as
Carlson and Land in
ref.\cite{L&C}) consider the limiting case $k_5=0$.
The CP violation in both cases is expressed by
$ \mu_{3},\mu_{4}$ and $k_{5} $ .

As usual, we express  $\Phi_{1},\Phi_{2}$ in terms of real
fields with the proper normalization of the kinetic term

\begin{equation}
\Phi_{1} = \frac{1}{\sqrt{2}} \left( \begin{array}{c}
\chi_{1}+\imath\eta_{1}\\
\phi_{1} + \imath\psi_{1} \end{array} \right) ,\qquad
\Phi_{2} =
\frac{1}{\sqrt{2}}\left( \begin{array}{c} \chi_{2} +
\imath\eta_{2}\\ \phi_{2} + \imath\psi_{2}\end{array}
\right) \end{equation}

As in \cite{L&C} we adjust the phases of the Higgs doublets
such that the only fields getting a vacuum expectation value
are $\phi_{1},\phi_{2},
\psi_{2}$.

To find the temperature dependent contribution to the potential
we proceed in
a
standard way \cite{Carrington},\cite{AndersonHall}. We only
take into
account lowest order effects and since only the top quark has
an appreciable
Yukawa coupling $\lambda_t$ we ignore the effects of the other
quarks and
derive the temperature dependent part of the effective potential
as

\begin{equation}
\frac{T^{2}}{2} \lbrack \alpha_{1}\phi_{1}^{2} +
\alpha_{2}(\phi_{2}^{2}+\psi_{2}^{2}) \rbrack
\end{equation}

where

\begin{equation}\begin{array}{lcl}
\alpha_{1} & = & 2 k_{1} + \frac{2}{3} k_{3} +
\frac{1}{3} k_{4} +\frac{3}{16}
g_{w}^{2} + \frac{1}{2}\lambda_{t}^{2}\\
 & & \\
\alpha_{2} & = & 2 k_{2} + \frac{2}{3} k_{3} +
\frac{1}{3} k_{4} +\frac{3}{16}
g_{w}^{2}
\end{array}\end{equation}

$\lambda_t$ is related to the top quark mass  by
$m_{top}^{2}=\lambda_{t} v^{2}$.
For our calculations we use a top mass $m_{t}=150$ GeV.
The W-Boson and top quark part of the $\alpha_{i}$'s is
calculated
using the standard procedure 

\begin{eqnarray}\label{Dis}
\alpha_{i} = 2 D_{i} , \qquad D_{1} = \frac{3 m_{w}^{2} +
2 m_{t}^{2}}{8
 v^{2}} ,\quad D_{2} = \frac{3 m_{w}^{2}}{8 v^{2}} \\
\alpha_{1} = \frac{3}{16} g^{2} + \frac{1}{2}\lambda_{t}^2 ,
\quad 
\alpha_{2} = \frac{3}{16} g^{2}
\end{eqnarray}

These values differ from those given in reference \cite{L&C}
by factors of $2$,
and therefore numerical results are substantially different
from those of reference \cite{L&C}.

For simplicity, we restrict our analysis from now on to
two-Higgs theories without explicit CP
violation in the Higgs sector, i.e. we set $k_{4} = k_{5} = 0$
and $\mu_{3} = 
\mu_{4} = 0 $. The dynamics of the phase
transition will not be affected substantially by small
non-vanishing couplings of this kind. This also implies that
the VEV's of $\Phi_i$ can be taken to be real, i.e.
$<\psi_2 >$ vanishes.

In this particular range of parameters the high temperature
effective potential is found to be

\begin{equation}
 V ( \phi_{1},\phi_{2},T) =  k_{1}\phi_{1}^{4} + k_{2}
\phi_{2}^{4} +
k_{3} \phi_{1}^{2}\phi_{2}^{2} -2 \mu_{1}^{2}\phi_{1}^{2} -
2 \mu_{2}^{2}\phi_{2}^{2}+\frac{T^{2}}{2}\lbrack \alpha_{1}
\phi_{1}^{2} + \alpha_{2}\phi_{2}^{2}\rbrack
\end{equation}

Using this potential we get for the temperature dependent
VEV's $v_{1}(T)^{2} =( 4 \mu_{1}^{2} -
\alpha_{1} T^{2})/4 k_{1}$ and $ v_{2}(T)^{2} = (4
\mu_{2}^{2} 
- \alpha_{2} T^{2} )/4 k_{2}$.
{}From the second derivative of the potential we
read off the zero temperature Higgs masses as
$m_{1}^{2} = 8 \mu_{1}^{2}$
and $ m_{2}^{2} = 2 k_{3} v^{2} -4 \mu_{2}^{2}$.

The critical temperatures $T_{c_2},T_{c_1}$ are defined as
those where the
symmetric (i.e. $\phi_i=0$) minimum becomes unstable into
the $\phi_2$
respectively $\phi_1$ direction. 
\begin{equation}
T_{c_1} = \sqrt{\frac{4 \mu_{1}^{2}}{\alpha_{1}}} ,
\qquad T_{c_2} =
\sqrt{\frac{4 \mu_{2}^{2}}{\alpha_{2}}}
\end{equation}

In order to have a two stage phase transition as proposed in
\cite{L&C} we
require $T_{c_1} < T_{c_2}$, such that the first stage of the
transition leads
to a minimum located at $(\phi_{1} = 0,\phi_{2} = v_{2}(T) )$.
Then at $
T_{c_1} $ the second minimum evolves which lies in the
$\phi_{1}$ direction.
As the temperature decreases the depths of the two minima
become
equal. This 
defines another temperature below which the
minimum in the $ \phi_{1} $
direction is lower than the minimum in $ \phi_{2} $ direction,
and critical bubbles may occur. This critical temperature is
given by

\begin{equation}
T_{c}^{2} = 4 \frac{\sqrt{k_{2}}\mu_{1}^{2} -
\sqrt{k_{1}}\mu_{2}^{2} }{
\sqrt{k_{2}}\alpha_{1} -\sqrt{k_{1}}\alpha_{2}}
\;.
\end{equation}

A further extremum is located at

\begin{equation}
\phi_{1}^{2}  =  \frac{2 k_{2} b_{1} - k_{3} b_{2}}{8
k_{1} k_{2}
-2 k_{3}^{2}} ,\qquad
\phi_{2}^{2}  =  \frac{2 k_{1} b_{2} - k_{3} b_{1}}{8
k_{1} k_{2}
-2 k_{3}^{2}}
\end{equation}

where $b_{1,2} = 4 \mu_{1,2} - \alpha_{1,2} T^{2} $.  This is
a saddle point separating the two minima. For certain
combinations
of the coupling constants we find that the saddle point moves
closer to the $\phi_{2}$
minimum as the temperature is falling below $T_{c}$ and merges
with it at the
roll-over temperature $T_{0}$. Now the
minimum in $\phi_{2}$ direction is not even classically stable.
The roll over temperature is

\begin{equation}
T_{0}^{2} = 4 \frac{2 k_{2} \mu_{1}^2 -k_{3} \mu_{2}^{2}}{2 k_{2}
\alpha_{1}
-k_{3} \alpha_{2}}
\end{equation}

The actual transition will take place in the temperature intervall
between the
tunneling and the roll-over temperature. In order to decide whether
the
transition is completed before roll-over takes place, we have to
study the
critical bubble solutions to get the free energy.

The existence of a roll-over temperature is not
necessary for a phase transition to
occur. In some parameter range, tunneling and bubble formation occur
close
to the critical temperature although the second minimum is preserved
for all
temperatures.

We require the following hierarchy for the above defined
temperatures in
order to have the desired two stage phase transition : 

\begin{equation}
T_{c_2} > T_{c_1} > T_{c} > 0
\end{equation}

The existence of $T_{c} > 0$ ensures that
$|V(v_{1},0)|>|V(0, v_{2})|$
holds
below $T_{c}$, i.e. the $\phi_1$ minimum is the true one.
Furthermore, we find that 
the
conditions $T_{c_2} > T_{c_1}$ and $T_{c_1} > T_{c}$ happen to
lead to the same
restrictions
on the free parameters of the theory, i.e.
 $m_{1},m_{2}$,
the two Higgs masses, and $k_{2},k_{3}$. Thus we get an upper
and a lower bound
for $k_{2}$ as a function of the other three parameters.
The upper bound corresponds to $T_{c_1} = T_{c_2}$, the lower
one to $T_c = 0$.
\section{Wash-out inside the bubble}
\setcounter{equation}{0}
The upper bound on $k_2$ is sharpened by taking into account
the wash-out behaviour. 
We require sufficient suppression
of baryon number violating sphaleron transitions in the
broken phase in order to maintain the baryon
asymmetry assumed to be generated during the second stage of
the phase transition, or before. This defines the wash-out
temperature $T_{wo}$ according to the followingline of
arguments.

We define the three-dimensional effective coupling $g_{3}$,
which is the relevant
temperature dependent expansion parameter for the effective
potential 
\cite{H&K&S}, as

\begin{equation}
g_{3}^{2} = \frac{g T }{v_{1}(T)}
\end{equation}

As pointed out in ref. \cite{Peccei} the sphaleron rate in
two-Higgs models
with positive couplings only, can fairly accurately be
estimated by an
effective coupling and the coresponding one-Higgs
sphaleron rate. In the case of a two stage phase transition
we have two kinds
of sphalerons, one when the $\phi_2$ minimum is the absolute
minimum (at $T>T_c$) and
the other, when the $\phi_1$ minimum plays this role
( at $T<T_c$).
Taking the boundary conditions for the Higgs fields of the
sphaleron
configuration and the positive definiteness of
$V(\phi_1,\phi_2)-V(min)$ into
account, we find that there is a sphaleron minimum of the
true vacuum only involving the $\phi_1$ field, whereas in
the intermediate phase only the $\phi_2$ field is
responsible. Thus, as long as we neglect CP violation,
the condition for a baryon number freeze-out inside the
true vacuum bubble reads

\begin{equation}
\frac{E_{Sph}(k_{1},T)}{T} = \frac{2 \pi}{g_{3}^{2}(T)}\; A
\leq 45
\end{equation}

where the function $A$ is of the order of 3.5, somewhat
depending on $4 k_{1}/g^2$ but only sligthly varying with T
( see Fig.2 of ref.\cite{H&K&S}).
The number $45$ arises from the comparison with the expansion
rate of the universe.

The above relation allows us to define the wash-out temperature,
below which the wash-out constraint is fullfilled: 

\begin{equation}
T_{wo} = \frac{E_{Sph}( k_{1}, 0)}{ \sqrt{45^{2} +2 \alpha_{1}
\lbrace E_{Sph}(
k_{1}, 0) /m_{1} \rbrace^{2}  } }
\end{equation}

We  define the nucleation temperature $T_{N}$ as the temperature
where the probability of
finding a critical bubble in the causal horizon of the universe
is equal to one. This temperature defines the start of the phase
transition with bubble expansion. We find $T_{N}$
(see \cite{AndersonHall}) by calculating the bubble solution and
the corresponding free energy numerically.

Since we are dealing with a two point boundary value problem in
two dimensions the numerical treatment turns out to be rather
involved.
Our numerical method consists of two steps. In the first step we
calculate an approximate bounce configuration. We use an adapted
'shooting' method and do the
necessary integrations of the equations of motion with the help
of the
Boerlisch-Stoer method. Then, we fit the approximate bounce
solution to the asymptotic behaviour of the fields.

In the second step we use a discrete set of points to describe
the
solution given by the first step. For this set of points we
minimize the sum
of the absolute value of the error of the equations of motion
in each point by using multidimensional minimisation
techniques.
The discussion of the numerical methods and the calculations
of general bounce solutions will be subject to a subsequent
publication \cite{Hammerschmitt&K&S}.

The nucleation temperature $T_{N}$ must be below $T_{wo}$.
For an optimal model
it should be close to $T_{wo}$ since otherwise baryon number
violation in the
intermediate phase is more strongly suppressed as well, and
a smaller baryon asymmetry is generated.

One can also define a temperature $T_{\Omega}$ where the phase
transition is
completed, and all space is converted to the true vacuum.
$T_{\Omega}$ is usually very close to $T_{N}$.

In order to clarify the meaning of all these temperatures
we discuss one
example. We consider the Higgs masses $m_1 = 70$ GeV and
$m_2 = 200 $ GeV, and choose $k_{3} = 0.425$. The various
critical temperatures are presented in
Figure 1 as function of $k_2$ for a fixed $k_{3}$.

\begin{figure}[h]
   \unitlength1cm
   \begin{picture}(14,8.3)
    \put(4.3,6.2){\small $T_{c_2}$}
    \put(3,4.5){\small $T_{c_1}$}
    \put(3,3.4){\small $T_{wo}$}
    \put(3.5,2){\small $T_{c}$}
    \put(7.5,2){\small $T_{N}$}
    \put(9.6,2){\small $T_{ro}$}
    \epsfxsize=11cm
    \epsfysize=7cm
    \put(1.5,0.2){\epsffile{Fig1.eps}}
    \put(7,0.2){\small $k_{2}$}
    \put(4.1,7.3){\small $k_{3}=0.425$, $m_{1}=70$,
$m_{2}=200$ [GeV]}
    \put(1,3.4){\rotate[l]{\small $T_{i}$ [GeV]}}
   \end{picture}
   \caption{Critical temperatures in the allowed
$k_{2}$ range}
\end{figure}

$T_{c_{1}},T_{wo}$ do not depend
on $k_{2}$. The point where the three critical temperatures
intersect gives
one upper bound on $k_2$, shown in Fig.2 as the upper curve.
The point where
the wash-out temperature and the nucleation temperature
$T_{N}$
meet, gives the true upper bound on $k_2$ if we
require baryon number violating processes being out of
equilibrium in the low-temperature phase.

At the crossing of $T_{wo}$ and  $T_{N}$, the two temperatures
$T_{\Omega}$ and $T_c$ are still very close together.
Therefore, for practical purposes, $T_c$ can be used to set
the upper bound on $k_2$.

In Fig.2 we display the accessible parameter space for a given
set of Higgs masses taking into account the above limits for
given $k_3$.

The lower bound on $k_2$ (Fig. 2) has been obtained from
requiring $T_c > 0$,
i.e. the minimum in $\phi_1$-direction being the true vacuum.
A stricter lower bound follows from $T_N >0$ ( i.e.
the phase transition should actually have taken place).

Since we want to maximize baryon number violation in the
intermediate phase the relevant $k_2$ range is close to
the upper bound.

The stronger upper bound comes from the wash-out condition.
It is remarkable
that the three curves intersect in two points (the other
located at $k_2 =
53.62$, $k_3 = 1.80$ outside the range of Fig.2) that give
a lower and upper
limit for the coupling constants $k_{2},k_{3}$. The allowed
parameter range is the narrow strip between the two
solid lines.

\begin{figure}[t]
  \unitlength1cm
  \begin{picture}(14,7.8)
    \put(8.3,2){\small $T_{c}>0$}
    \put(9.2,4){\small $T_{N}>0$}
    \put(8,6){\small $T_{wo}< T_{c}$}
    \put(5,3.5){\small $T_{c_1}< T_{c_2}$}
    \put(8.8,5.9){\vector(1,-1){0.88}}
    \epsfxsize=11cm
    \epsfysize=7cm
    \put(1.5,0.2){\epsffile{Fig2.eps}}
    \put(1,3.9){\rotate[l]{\small $k_{2}$}}
    \put(7,0.1){\small $k_3$}
    \put(5.1,7.4){\small $m_{1}=70$, $m_{2}=200$ [GeV]}
  \end{picture}
  \caption{The allowed parameter range}
\end{figure}

The range of $k_2$ values displayed in Fig.2 is not
realistic. Renormalized
scalar self-couplings cannot be chosen arbitrarily large.
There will be
triviality bounds. For the $O(4)$ theory with one Higgs
vector this has been
studied in detail ( see, e.g. ref. \cite{Luescher}). For
the standard model, extended by a second Higgs doublet,
such an analysis is not
available. The couplings $k_i$ should not exceed some
number of order one,
however. Therefore, only the lower-left corner of Fig.2
is relevant.

This still leaves open a large region in the $m_1-m_2$
plane, however.

Restrictions on the masses arise from triviality bounds
on $k_{3}$, not
$k_{2}$. The smallest possible $k_{3}$ for a given set
of Higgs masses is given by

\begin{eqnarray}
 k_{3}^{c} & = & 3 D_{1} +\frac{m_{1}^{2} +
    8 m_{2}^{2}}{8 v^{2}} +
   \nonumber\\
    & & \sqrt{
     \frac{ ( m_{1}^{2}-4m_{2}^{2} )^{2}}{64 v^{4}} +
\frac{3D_{1}(m_{1}^{2}+4m_{2}^{2})}{4 v^{2}}+9D_{1}^{2}-
   \frac{3 D_{2} m_{1}}{2 v^{2}}  }
\end{eqnarray}

 where the $D_{i}$ are defined in eq. (\ref{Dis}). Solutions
with the correct
 hierarchy pattern exist as long as $k_{3}^{c}$ does not
exceed the triviality
 bound $\bar{k}$. In Fig.3 we show the allowed region in the
$m_{1},m_{2}$ plane for different $\bar{k}$'s.

 \begin{figure}[t]
  \unitlength1cm
  \begin{picture}(14,8)
    \put(8,2){\small $\bar{k}= 0.25$}
    \put(5.8,4){\small $\bar{k}= 0.5$}
    \put(3.6,6){\small $\bar{k}= 1$}
    \epsfxsize=11cm
    \epsfysize=7cm
    \put(1.5,0.2){\epsffile{Fig3.eps}}
    \put(6.7,0.1){\small $m_1$ [GeV]}
    \put(1,3.2){\rotate[l]{\small $m_{2}$ [GeV]}}
  \end{picture}
  \caption{Critical mass $m_{2}$ as a function of $m_{1}$
for different triviality bounds}
\end{figure}

These curves give an upper bound for the second Higgs mass
for given triviality bounds $\bar{k}$.

\section{Wash-out in the intermediate phase}
\setcounter{equation}{0}

If we now insist on generating the baryon asymmetry during
the phase
transition
(see, e.g., the scenarios proposed by
\cite{CohenKaplanNelson},\cite{CoKaNe2} or,
\cite{Shaposhnikov}),
we need strong B violation in the false vacuum phase, in order
to wash out any accumulated anti-baryon number in front of, or
perhaps inside the bubble wall. This is a requirement in
conflict with avoiding the baryon
wash-out in the final phase. Lowering the transition temperature
will also reduce the sphaleron rate in the metastable
intermediate phase. The generated
baryon asymmetry will be proportional to this rate, however,
quite independently from details of the mechanism.

The sphaleron rate in the intermediate phase is difficult to
estimate because we have to handle strong coupling problems
both regarding $k_2$, and the
effective gauge coupling $g^2_3(T)=gT/v_2(T)$.
Actually, increasing the sphaleron rate immediately implies
reducing $v_2(T)$, and thereby increasing the effective gauge
coupling. Nevertheless, the Higgs
field still provides an infrared cut-off, and therefore the
situation is better under control than for the high
temperature phase with unbroken symmetry.

Non-perturbative methods for estimating
sphaleron rates at stronger coupling are being developed
\cite{krimich,bielef} but no results for realistic theories
are available yet. Therefore, we will simply assume that the
quasiclassical expression for the sphaleron rate would
still be applicable in the domain of stronger coupling
\begin{equation}\label{Unser}
\Gamma = \frac{\omega_{-}}{2 \pi}{\cal N}_{tr} ({
\cal N} V)_{rot}
\left( \frac{\alpha_{w} T}{4 \pi}  \right)^{3}
\alpha_{3}^{-6} e^{\frac{E_{Sph} }{T}} \kappa
\end{equation}
where we follow the notation of \cite{McLerr2},\cite{H&K&S}.

We compare the wash-out rate of the intermediate phase and
the true vacuum phase to find out where the strongest
wash-out in the intermediate phase can be
achieved. This ratio is governed by $E_{Sph}(k_{2},T)$ and
$ E_{Sph}(k_{1},T)$, whose quotient we take as a measure.

\begin{figure}[h]
\unitlength1cm
\begin{picture}(14,7.5)
  \put(7.5,6){\small $\bar{k}=0.25$}
  \put(8,4){\small $\bar{k}=2$}
  \put(11,5.6){\small $\bar{k}=0.5$}
  \epsfxsize=12cm
  \epsfysize=7cm
  \put(1.5,0.2){\epsffile{Fig4.eps}}
  \put(6.7,0.1){\small $m_1$ [GeV]}
  \put(1,1.5){\rotate[l]{
\small $E_{Sph}(k_{2},T_{wo})/ E_{Sph}(k_{1},T_{wo})$ }}
\end{picture}
\caption{Minimal ratio
$E_{Sph}(k_{2},T_{wo})/ E_{Sph}(k_{1},T_{wo})$ for
different triviality bounds $\bar{k}$ as a function of the
Higgs mass}
\end{figure}

In order to find the minimal ratio we will take the $k_{2}$
that lies on the wash-out curve of Fig.2 where the wash-out
bound is just fullfilled, i.e. we are at the maximal
temperature for a given $k_{3}$.

We find that the minimal ratio
$E_{Sph}(k_{2},T_{wo})/ E_{Sph}(k_{1},T_{wo})$
solely depends on $m_{1}$. This is due to the fact
that the only expressions involving $m_{2}$ are
$v_{2}(T_{wo})$ and $T_{wo}$.
But since their quotient is independent of $m_{2}$, the ratio
itself is independent of the second Higgs mass. In figure
$4$ we show the dependence of the calculated ratio on
$m_{1}$ for various triviality bounds on $k_{2}$.

In this case $k_{2}$ hits the triviality bound before
$k_{3}$ ( compare figure 2). We find the best wash-out
behaviour in the intermediate phase for small $m_{1}$.

In order to decide whether this wash-out rate is strong
enough, $\Gamma$ of eq.(\ref{Unser}), applied to the
intermediate phase,should be compared with the expression
adopted on dimensional grounds for the symmetric phase.

\begin{equation}\label{1Higgs}
\Gamma_{o} = \kappa_{o} ( \alpha_{w} T )^{4}\;,
\end{equation}

where $\kappa_{o}$ is a dimensionless constant asumed to
be of order $1$.
Compared to these calculations we will predict a
baryon asymmetry that is smaller by the ratio of the
sphaleron rate of eq. (\ref{Unser}) to the rate
eq.(\ref{1Higgs}).

The ratio of $\Gamma / \Gamma_{o}$ shows how close we can
get with our models to the required wash-out rate $\Gamma_{o}$.

It is shown in figure 5 that at smaller Higgs masses close to
the experimental limit we loose about $4-5$ orders of magnitude
in the baryon asymmetry compared to estimates
starting from the symmetric phase. At larger Higgs masses the
suppression increases dramatically.

\begin{figure}[h]
\unitlength1cm
\begin{picture}(14,7)
  \epsfxsize=12cm
  \epsfysize=6.8cm
  \put(1.5,0.2){\epsffile{Fig5.eps}}
  \put(6.7,0.1){\small $m_1$ [GeV]}
  \put(1,2.7){\rotate[l]{
\small $-\log_{10}(\Gamma / \Gamma_{o})$ }}
\end{picture}
\caption{Optimal $\Gamma / \Gamma_{o}$ over $m_1$}
\end{figure}

{}From this point of view it seems very difficult to create the
baryon asymmetry of the universe at the phase transition,
despite the strong coupling uncertainties.

\section{Conclusion}
\setcounter{equation}{0}
The viability of the class of two-Higgs models considered
for explaining the baryon asymmetry of the universe through
the electroweak phase transition depends crucially on the
sphaleron rate in the intermediate phase. If this rate
is much smaller than the assumed rate
of B+L violation in the high temperature symmetric phase,
no significant baryon
asymmetry can be generated. The problem of washing-out a
baryon asymmetry in the final Higgs
phase can be easily avoided, however, without introducing
upper bounds on the Higgs mass conflicting with experimental
lower bounds.

Estimating the sphaleron rate in the intermediate phase is
a non-perturbative strong coupling problem, and therefore
no reliable results are available. It does not seem plausible,
however, that this rate could be high enough for baryon number
generation.

\end{document}